\documentclass[12pt, a4paper]{amsart}
\usepackage{graphicx}
\usepackage{amssymb,latexsym}
\usepackage[cp1251]{inputenc}
\usepackage[english]{babel}
\usepackage{textcomp}

\theoremstyle{plain}
\newtheorem{theorem}{Theorem}
\newtheorem{proposition}{Proposition}
\newtheorem{lemma}{Lemma}
\theoremstyle{remark}
\newtheorem{remark}{Remark}

\numberwithin{theorem}{section}
\numberwithin{proposition}{section}
\numberwithin{lemma}{section}
\numberwithin{remark}{section}
\numberwithin{equation}{section}

\newcommand{\N}{\mathbb N}

\newcommand{\R}{\mathbb R}

\newcommand{\entropy}{\mathcal{S}}

\newcommand{\expectation}{\mathbb{E}}

\newcommand{\gammaclass}{\mathcal{G}}

\renewcommand{\P}{\mathbb{P}}
\newcommand{\G}{\mathfrak{G}}
\newcommand{\F}{\mathcal{F}}
\newcommand{\K}{K}

\newcommand{\cav}{\mathrm{Cav\,}}

\newcommand{\val}{\mathrm{val}}
\newcommand{\V}{V}
\newcommand{\M}{\mathcal{M}}
\newcommand{\one}{\mathbf{1}}

\newcommand{\I}{I}
\newcommand{\J}{J}
\newcommand{\supp}{\mathrm{supp}\,}
\newcommand{\err}{\mathrm{err}}
\renewcommand{\entropy}{\mathrm{S}}
\newcommand{\Z}{\mathrm{Z}}
\newcommand{\tv}{\mathrm{TV}}
\newcommand{\nr}{\mathrm{NR}}
\newcommand{\A}{\mathrm{A}}

\renewcommand{\L}{\mathrm{L}}

\begin{document}

\title[]{Repeated games of incomplete information with large sets of states}


\vskip 1cm
\maketitle
\begin{center}
Fedor Sandomirskiy\footnote[1]{Chebyshev Laboratory
(Faculty of Mathematics and Mechanics, St.Petersburg State University)}\footnote[2]
{Department of Mathematical Physics (Faculty of Physics, St.Petersburg State University).}
\vskip 0.2cm
e-mail: sandomirski@yandex.ru\\
\end{center}
\begin{abstract}
The famous theorem of R.~Aumann and M.~Maschler states that
the sequence of values of an $N$-stage zero-sum game $\Gamma_N(\rho)$ with incomplete information on one side and prior distribution $\rho$ converges as $N\to\infty$, and the error term
$\err[\Gamma_N(\rho)]=\val[\Gamma_N(\rho)]-\lim_{M\to\infty}\val[\Gamma_{M}(\rho)]$ is bounded by $C N^{-\frac{1}{2}}$ if the set of states $\K$ is finite.
 
The paper deals with the case of infinite $\K$. 
It turns out that for countably-supported prior distribution $\rho$ with heavy tails the
error term can be of the order of $N^{\alpha}$ with $\alpha\in \left(-\frac{1}{2},0\right)$, i.e., the convergence can be anomalously slow. The maximal possible $\alpha$ for a given $\rho$ 
is determined in terms of entropy-like family of functionals.

Our approach is based on the well-known connection  
between the behavior of the maximal variation of measure-valued  martingales and
asymptotic properties of repeated games with incomplete information.
\end{abstract}
\section{Introduction}\label{sect_intro}
Repeated zero-sum games with
incomplete information on one side were introduced
by R.~Aumann and M.~Maschler (for a comprehensive presentation of the theory of repeated games we refer to the books~\cite{A-M}~and~\cite{Bigbook}). 
In such a game $\Gamma_N(\rho)$ two players are involved in a multistage repeated interaction, but only
Player~1 is completely informed of its properties,
and Player~2 has an uncertainty about the actual payoffs that
depend on a random $\rho$-distributed state $k\in\K$ chosen by Nature before the game starts; $k$ is told to Player~1 but not to Player~2, who knows only $\rho$.  
The total payoff received by Player~1 from Player~2 at the end of the game is
the expected arithmetic mean of one-stage gains. 
Both players are rational.

As it was shown by R.~Aumann and M.~Maschler, 
the minimax values of $\Gamma_N(\rho)$ converge as $N\to\infty$,
and the error term $\err[\Gamma_N(\rho)]=\val[\Gamma_N(\rho)]-\lim_{M\to\infty}\val[\Gamma_M(\rho)]$
is non-negative and is bounded from above by $\frac{C}{N}\Psi_N(\rho)$, where $\Psi_N(\rho)$ is the value
of the following optimization problem: to maximize the expected sum of distances between consecutive values up to time $N$ over all martingales $(\mu_n)_{n\geq 0}$ taking values in probability distributions on $\K$ with $\mu_0=\rho$ (the total variation distance is used).
The quantity $\Psi_N(\rho)$ is called
the maximal variation and represents the maximal variability of beliefs during
the process of Bayesian learning with prior $\rho$.
 R.~Aumann and M.~Maschler proved
that, if $\K$ is finite, then $\Psi_N(\rho)$ is less than constant times $\sqrt{N}$, and, therefore,
the error term can not decrease slower than $1/\sqrt{N}$. 
In~\cite{Zamir_sqrt_is_precise}, S.~Zamir showed the existence
of games with the error term
of the order of $1/\sqrt{N}$.

In the paper~\cite{MZBothSides} of J.-F.~Mertens~and~S.~Zamir,  the following inequality was derived
$\Psi_N(\rho)\leq\sqrt{N}\sum_{k\in\K}\sqrt{\rho(\{k\})(1-\rho(\{k\}))}$.
Hence, for countably infinite $\K$ with $\rho$ having not too heavy tails the maximal variation also can not grow faster than $\sqrt{N}$.
In~\cite{Neyman}, A.~Neyman
estimated the maximal variation by Shannon's entropy
$\Psi_N(\rho)\leq\sqrt{2 N {\entropy}(\rho)}$, where $\entropy(	\rho)=-\sum_{k\in{\K}}\rho(\{k\})\ln(\rho(\{k\}))$.
For countably infinite $\K$ this result significantly extends
the class of $\rho$ with $\sqrt{N}$-rate of growth of the maximal variation and 
with corresponding upper bound on the error
term. However it was unknown what happens if the tails of $\rho$ are so heavy
that $\entropy(\rho)=\infty$. The question
about the exact class of $\rho$ with $\sqrt{N}$-behavior of the maximal variation
was mentioned in~\cite{Neyman}. 

In this paper countably-supported prior distributions with heavy tails are considered. We  describe asymptotic behavior of the maximal variation and the slowest possible rate of decreasing of the error term. 
It turns out that for heavy-tailed $\rho$ anomalous behaviors are possible. 
Let $\alpha_\Psi(\rho)$ be such $\alpha$ that $\Psi_N(\rho)$ grows like $N^{\alpha}$, and
let $\alpha_\gammaclass(\rho)$ be the maximal $\alpha$ such that there is a game $\Gamma$ with $\err[\Gamma_N(\rho)]$ decreasing like $N^{\alpha}$ (the rigorous definitions of $\alpha_\Psi$ and $\alpha_\gammaclass$ are given in Section~\ref{sect_results}).
We set $$\Z_\varepsilon(\rho)=\sum_{k\in\K}\rho(\{k\})
\left[\ln\left(\frac{1}{\rho(\{k\})}\right)\right]^{\frac{1}{2}-\varepsilon}.$$
Our central result is the following identity $$\alpha_\Psi(\rho)=\alpha_\gammaclass(\rho)+1=\frac{1}{2}+\varepsilon^*(\rho),\quad\mbox{where} \ \, \varepsilon^*(\rho)=\inf\left\{\varepsilon\in[0,1/2]\mid \Z_\varepsilon(\rho)<\infty\right\}.$$
In particular, the class of $\rho$ with $\sqrt{N}$-behavior prescribed
by Neyman's condition $\entropy(\rho)<\infty$ can be extended again, and the exact
class is given by the condition of $\Z_0(\rho)$ finiteness. 
We also discuss the case of uncountable state space $\K$.

The results of this paper were announced in~\cite{Fedor_doklady}.
\section{Repeated games with incomplete information: main definitions}
\label{sect_definitions}
Here we describe an $N$-stage zero-sum repeated game $\Gamma_N(\rho)$ with incomplete information on the side of Player~2. This game
is given with a 4-tuple $\Gamma=(\K,\I,\J,\A)$,
a number of repetitions $N\in\N$, and a prior distribution $\rho\in\Delta(\K)$.
Here $\K$ is a set of states; $\I$ and $\J$ are sets of actions of Player~1 and Player~2,
respectively; $A:\K\times\I\times\J\to\R$ is a one-stage payoff function; $\Delta(\K)$ denotes 
the set of all probability distributions on $\K$.

The game is played as follows. 
Before the beginning of the game Nature picks a state $k\in\K$ at random  
with distribution $\rho$ and tells $k$ to Player~1. Player~2 knows only $\rho$.
Then at each stage $n=1,2,...N$
players simultaneously select their actions $i_n\in \I$ and  $j_n\in \J$ using the information they have at this stage, and these actions are publicly announced before the stage
$n+1$. A behavioral strategy $\sigma$ of Player~1 is a sequence of maps $\sigma_n:\ \K\times (\I\times \J)^{n-1}\to\Delta(\I)$. Player~1 randomizes his action $i_n$ according to $\sigma_{n}$ given $k$ and
a history $(i_1,j_1,...i_{n-1},j_{n-1})$ observed. Behavioral strategy $\tau$ of Player~2 consists of $\tau_n:\ (\I\times \J)^{n-1}\to\Delta(\J)$.
The prior distribution with strategies $\sigma$ and $\tau$ generate the probability
measure $\P_{\rho,\sigma,\tau}$ on $\K\times (\I\times\J)^N$. 
After the last stage Player~2 pays
\begin{equation}\label{eq_g_N_definition}
g_N(\rho,\sigma,\tau)=\frac{1}{N}\expectation_{\rho,\sigma,\tau}\left(\sum_{n=1}^N \A_{i_n,j_n}^k\right)
\end{equation}
to Player~1 (expectation is with respect to $\P_{\rho,\sigma,\tau}$).

Hence, players have completely opposite goals. 
Player~1 aims to maximize
$g_N(\rho,\sigma,\tau)$, and Player~2
wants to minimize it.
The lower and upper values 
are given by 
$\underline\val[\Gamma_N(\rho)]=\sup_\sigma\inf_\tau g_N(\rho,\sigma,\tau)$ and 
$\overline\val[\Gamma_N(\rho)]=\inf_\tau \sup_\sigma g_N(\rho,\sigma,\tau)$, respectively.
If these values coincide, the game has a value $\val=\underline\val=\overline\val$.
For finite $\I$, $\J$, and $\K$ the existence of the value follows from von~Neumann's minimax theorem.

The non-revealing game $\Gamma_1^{{\nr}}(\rho)$ is an auxiliary version of the 
one-stage game $\Gamma_1(\rho)$,
where Player~1 forgets $k$, i.e., the sets of strategies of Player~1 and Player~2 can be identified
with $\Delta(\I)$ and $\Delta(\J)$, respectively. 

We denote by $\gammaclass(\K)$ the class of all 4-tuples $\Gamma=(\K,\I,\J,\A)$ such that:
\begin{itemize}
\item the sets of actions $\I$ and $\J$ are countable;
\item the norm $\|\A\|_\infty=\sup_{k,i,j}|\A_{i,j}^k|$ is finite;
\item the games
$\Gamma_N(\rho)$ and $\Gamma_1^{\nr}(\rho)$ have values for any $\rho\in\Delta(\K)$ and $N\in\N$.
\end{itemize}
It is natural to consider the class $\gammaclass(\K)$ if we are going to deal
with infinite $\K$.
The first assumption allows us to avoid measurability issues
arising for uncountable $\I,\J$.
The second assumption ensures that
the anomalous asymptotic effects for infinite $\K$ are caused by ``infinite lack of knowledge''
on the side of Player~2 and not by unbounded payoffs. The third assumption 
excludes some pathological situations.

\subsection{The maximal variation and its role}
\label{subsect_maximal_variation}
Let $\K$ be countable.
The sequence of measure-valued maps $\mu_n: \ \Omega\to\Delta(\K)$ 
defined on a filtered probability space $(\Omega,\F,(\F_n)_{n\geq 0},\P)$
is called an $(\F_n)_{n\geq 0}$-adapted $\Delta(\K)$-valued martingale
if the sequence $(\mu_n(B),\F_n)_{n\geq 0}$ forms a martingale for any $B\subset\K$. 

We denote by ${\M_{\Delta(\K)}}(\rho)$ the set of all $\Delta(\K)$-valued martingales $\mu=(\mu_n,\F_n)_{n\geq0}$
with non-random $\mu_0=\rho$. The probability space is not fixed, i.e., more precisely,
${\M_{\Delta(\K)}}(\rho)$ consists of pairs $\mu=\big((\mu_n,\F_n)_{n\geq0},\ (\Omega,\F,(\F_n)_{n\geq 0},\P)\big)$. For brevity we write $\mu\in{\M_{\Delta(\K)}}(\rho)$ thus
implicitly fixing the underlying probability space and we use $\P$ for
the underlying probability measure and $\expectation$ for the expectation with respect to $\P$.

The $N$-term variation of a $\Delta(\K)$-valued martingale $\mu$
is defined by
$$\V_N(\mu)=\expectation \left(\sum_{n=0}^{N-1}\|\mu_{n+1}-\mu_n\|_\tv\right),
$$
where $\|\Phi\|_\tv$ denotes the total variation of a signed measure $\Phi$. In other words,
$\V_N(\mu)=\sum_{n=0}^{N-1}\sum_{k\in\K}\expectation |\mu_{n+1}(\{k\})-\mu_n(\{k\})|$.
Taking supremum over all $\mu$ from ${\M_{\Delta(\K)}}(\rho)$ we get the maximal variation 
$$\Psi_N(\rho)=\sup_{\mu\in\M_{\Delta(\K)}(\rho)}\V_N(\mu).$$

The maximal variation is extremely important in asymptotic problems concerning repeated games with incomplete information.
The fundamental estimate on the value follows from the results of R.~Aumann and M.~Maschler. 
Denote by $u_\Gamma(\rho)$ the value of the non-revealing game $\Gamma_1^{\nr}(\rho)$.
Let $\cav[u_\Gamma]$ be the least concave majorant of $u_\Gamma$ as a function of $\rho$, i.e.,
$\cav[u_\Gamma](\rho)=\sup\sum_{m=1}^M \beta_m u_\Gamma(\rho_m)$, where
supremum taken is over all $M\in\N$, $\{\beta_m\}_{m=1,2,...M}\subset[0,1]$, and $\{\rho_m\}_{m=1,2,...M}\subset\Delta(\K)$ such that $\sum_{m=1}^M \beta_m=1$
and $\sum_{m=1}^M\beta_m\rho_m=\rho$.
\begin{theorem}[R.~Aumann and M.~Maschler]\label{th_cavu}
Suppose $\K$ is countable and $\Gamma\in\gammaclass(\K)$.
Denote $\val[\Gamma_N(\rho)]-\cav[u_\Gamma](\rho)$
by $\err[\Gamma_N(\rho)].$
Then the following two-sided estimate holds: 
\begin{equation}
\label{eq_error_term_bound}
0\leq \err[\Gamma_N(\rho)]\leq \frac{\|\A\|_\infty}{N}\Psi_N(\rho).
\end{equation}
\end{theorem}
Usually this result is formulated for finite $\K$, $\I$ and $\J$,
but the same proof (see~\cite{Bigbook}, Theorem~2.10 p.225) works in our case. 
\begin{remark}
Suppose Player~1 uses a behavioral strategy $\sigma$. Then Player~2 can compute
the conditional distribution $\rho_n$ of $k$ after observing the history of actions
$(i_1,j_1,...i_n,j_n)$.
The reason
why $\Delta(\K)$-valued martingales arise is that
the process $(\rho_n)_{n\geq 0}$ of Player~2 beliefs about $k$ belongs to $\M_{\Delta(\K)}(\rho)$. 
Note that more general functionals similar to the variation
also appear in the context of repeated games (see the paper~\cite{DeM2010} of B.~De~Meyer and the papers~\cite{CavU, Gensbittel2013_fresh} of F.~Gensbittel).
Because of the presence of other maximal variations it is more correct to call
the particular one defined above ``the maximal variation in the total variation norm''. For brevity we use the shorter notation.
\end{remark}

It is well known that for countable $\K$
and $\rho$ with not too heavy tails $\lim_{N\to\infty}\Psi_N(\rho)/N=0$ (see below).
Therefore, $\val[\Gamma_N(\rho)]$ converges to $\cav[u_\Gamma](\rho)$. This limiting value
incorporates the influence of constant strategic
advantages (or disadvantages) inherited from the non-revealing game, and the 
error term $\err[\Gamma_N(\rho)]$ 
reflects the impact of information asymmetry on the value. In particular, if $u_\Gamma\equiv 0$ (in this case all the strategic asymmetries in $\Gamma_N(\rho)$ are caused by incomplete information), then the error term can be regarded as the price of information.

In~\cite{MZBothSides}, J.-F.~Mertens~and~S.~Zamir showed that
\begin{equation}
\label{eq_Aumann_variation_estimate}
\Psi_N(\rho)\leq\sqrt{N}\Lambda(\rho),\ \ \mbox{where} \ \
\Lambda(\rho)=\sum_{k\in\K}\sqrt{\rho(\{k\})(1-\rho(\{k\}))}
\end{equation}
(the similar but weaker estimate was derived earlier by R.~Aumann and M.~Maschler).
Hence, the maximal variation can not grow faster than $\sqrt{N}$ if
$\Lambda(\rho)<\infty$ (i.e., if $\rho$ has not too heavy tails). 
For $\Gamma\in\gammaclass(\K)$ this implies $C/\sqrt{N}$ upper bound on the error term.
S.~Zamir) proved that 
the order of magnitude of this upper bound is sharp (see~\cite{Zamir_sqrt_is_precise}).
In other words, for any non-degenerate $\rho\in\Delta(\K)$ (i.e., not concentrated at one point) there is a $4$-tuple $\Gamma\in\gammaclass(\K)$
such that  $\err[\Gamma_N(\rho)]\geq 1/\sqrt{N}$ for any $N$.
Therefore, $1/\sqrt{N}$ is the slowest possible rate of the error term decreasing
over $\gammaclass(\K)$ for $\rho$ with $\Lambda(\rho)<\infty$.
In the paper~\cite{Neyman}, A.~Neyman extended the class of $\rho$ 
with $\sqrt{N}$-behavior of the maximal variation by obtaining the following estimate
in terms of Shannon's entropy
\begin{equation}
\label{eq_Neyman_entropy}
\Psi_N(\rho)\leq\sqrt{2 N {\entropy}(\rho)},\quad
\entropy(\rho)=\sum_{k\in{K}}\rho(\{k\})\ln\left(\frac{1}{\rho(\{k\})}\right),
\end{equation}
where $\ln x$ denotes the logarithm of $x$ to the base $e$.
Indeed, if $\K=\N$, then
$\Lambda(\rho)$ diverges for $\rho$ with $\rho(\{k\})\sim k^{-2}$ as $k\to\infty$, but Shannon's entropy remains
finite even if $\rho(\{k\})\sim k^{-1}(\ln k)^{-2-\varepsilon}$ with some $\varepsilon>0$.
\section{The results}\label{sect_results}
\subsection{Countable $\K$}\label{subsect_countable_K}
The exact class of $\rho\in\Delta(\K)$ with the maximal variation growing like $\sqrt{N}$ turns out to be wider than prescribed by the condition of entropy finiteness.
Consider a family of uncertainty measures of $\rho$
$$\Z_\varepsilon(\rho)=\sum_{k\in\K}\rho(\{k\})
\left[\ln\left(\frac{1}{\rho(\{k\})}\right)\right]^{\frac{1}{2}-\varepsilon},\quad  \varepsilon\leq{\frac{1}{2}}.$$
The Shannon entropy corresponds to $\varepsilon=-\frac{1}{2}$.
We will see that the sharp condition for $\sqrt{N}$-behavior to hold is finiteness
of $\Z_0(\rho)=\sum_{k\in\K}\rho(\{k\})
\sqrt{\ln\left(\frac{1}{\rho(\{k\})}\right)}$. 
The condition $\Z_0(\rho)<\infty$ is less restrictive than entropy finiteness; if $\K=\N$, it
holds for $\rho$ such that
$\rho(\{k\})\sim k^{-1}(\ln k)^{-\frac{3}{2}-\varepsilon}$ as $k\to\infty$ with some $\varepsilon>0$, but
entropy is infinite if $\varepsilon\leq\frac{1}{2}$.
Moreover, it turns out that beyond the class of $\rho$ with finite $\Z_0(\rho)$ 
the maximal
variation can grow anomalously fast (like $N^{\frac{1}{2}+\varepsilon}$ with some $\varepsilon\in(0,1/2)$), and the error term can decrease anomalously slowly.

Our main goal is to study the exponents
$$\alpha_\Psi(\rho)=\limsup_{N\to\infty}\frac{\ln\left(\Psi_N(\rho)\right)}{\ln N}\ \ \ \mbox{and} \ \ \ \alpha_\gammaclass(\rho)=\sup_{\Gamma\in\gammaclass({K})}
\limsup_{N\to\infty}\frac{\ln\left(\err[\Gamma_N(\rho)]\right)}{\ln N}$$
as functions of $\rho$.

The following theorem provides a one-parametric family of estimates on the maximal variation.
\begin{theorem}\label{th_maximal_variation_upper_estimate}
Suppose $\K$ is countable, and $\rho\in\Delta(\K)$. Then
for any $\varepsilon\in[0,1/2]$ 
\begin{equation}
\label{eq_estimate_variation_from_above}
\Psi_N(\rho)\leq c N^{\frac{1}{2}+\varepsilon}\Z_\varepsilon(\rho),
\end{equation}
where $c=\sqrt{2}\left(1+\frac{1}{2\ln 2}\right)<\sqrt{6}$.
\end{theorem}
The next theorem states that the estimate (\ref{eq_estimate_variation_from_above}) after minimizing over $\varepsilon$ gives the exact rate of growth of the maximal variation.
\begin{theorem}\label{th_maximal_variation_alpha_Psi}
If $\K$ is countable, and $\rho\in\Delta(\K)$ is non-degenerate, then
\begin{equation}
\label{eq_alpha_psi_statement}
\alpha_\Psi(\rho)=\frac{1}{2}+\varepsilon^*(\rho), \qquad \varepsilon^*(\rho)=\inf\left\{\varepsilon\in[0,1/2]\mid \Z_\varepsilon(\rho)<\infty\right\}.
\end{equation}
\end{theorem}
Theorem~\ref{th_maximal_variation_upper_estimate}~and~\ref{th_maximal_variation_alpha_Psi}
are proved in Sections~\ref{sect_from_above}~and~\ref{sect_anomalous}, respectively.

Note that $\Z_{\frac{1}{2}}(\rho)=1$ and, hence, $\varepsilon^*(\rho)$ is well-defined for any $\rho$. If tails of $\rho$
are so heavy that $\varepsilon^*(\rho)=\frac{1}{2}$, then a natural question about 
possibility of linear growth of $\Psi_N(\rho)$ arises. The negative answer is given in Section~\ref{sect_from_above},
where we show that $\frac{1}{N}\Psi_N(\rho)\to 0$  as $N\to\infty$
for any countably-supported $\rho$.  

From the results described in
Subsection~\ref{subsect_maximal_variation} it follows that for non-degenerate $\rho$ with $\entropy(\rho)<\infty$ we have $\alpha_\gammaclass(\rho)=\alpha_\Psi(\rho)-1=-1/2$.
The relation between $\alpha_\Psi$ and $\alpha_\gammaclass$ holds in general case. 
\begin{theorem}\label{th_alpha_Gamma}
If $\K$ is countable, and $\rho\in\Delta(\K)$ is non-degenerate, then
\begin{equation}
\label{eq_alpha_identity}
\alpha_\gammaclass(\rho)=-\frac{1}{2}+\varepsilon^*(\rho)=\alpha_\Psi(\rho)-1.
\end{equation}
\end{theorem}
This theorem is proved in Section~\ref{sect_game_with_anomalous} by  
constructing a game $\G_N(\rho)$ with anomalously slow
decreasing of the error term. 
\begin{remark}
If $\K$ is countable, the game $\G_N(\rho)$
has countably infinite sets of actions of both players. The infiniteness 
of action sets turns out to be crucial for anomalous behavior of the error term. Indeed, as it was announced in the paper~\cite{Neyman} of A.~Neyman, for finite $\I$ and $\J$
the error term is bounded from above by
$4\|\A\|_\infty\sqrt{2\#\I\#\J\ln 2}/{\sqrt{N}}$ (here $\# B$
denotes the cardinality of a set $B$). 
\end{remark}

\subsection{Uncountable $\K$}\label{subsect_results_cont_K}
If $\K$ is a ``good'' uncountable measurable space, then one can consider repeated
games with incomplete information and the maximal variation
 after obvious refinements of definitions.
\begin{remark}
A completely metrizable separable 
topological space $\K$ equipped with its Borel sigma-field is called a Polish space
(see \cite{Srivastava}, p.52). 
It is enough to keep in mind one of the following examples: countable sets with the discrete topology;
interval $[0;1]$ with the standard topology of the real line (or the real line itself).
These are the only Polish spaces up to Borel isomorphism, i.e., up to bijection preserving
the Borel structure (\cite{Srivastava}, Theorem~3.3.13 p.99). 
The set  $\Delta(\K)$ of Borel probability measures over a Polish space
$\K$ with the topology of weak convergence becomes Polish itself, and, hence, one can define $\Delta(\K)$-valued measurable maps.

In order to consider repeated games with Polish $\K$, $\I$ and $\J$ we assume that:
the one-stage 
payoff function $\A:\ \K\times\I\times\J\to\R$ is measurable; behavioral
strategies of Player~1 and Player~2 consist of measurable maps with values
in $\Delta(\I)$ and $\Delta(\J)$, respectively. For a Polish space $\K$ 
 a sequence of $\Delta(\K)$-valued 
random variables $(\mu_n)_{n\geq 0}$ is called a $\Delta(\K)$-valued martingale if  for any measurable
$B\subset\K$ the sequence $(\mu_n(B))_{n\geq 0}$ is a martingale. 
Up to this extension we define the maximal variation as before.
For countable $\K$ the new definitions are equivalent to the given above.
\end{remark} 

Note that
$\rho\in\Delta(\K)$ can be 
represented as the sum of the continuous component $\rho^c$ that has no atoms and the discrete component $\rho_d$ supported on a countable subset of 
$\K$. If $\rho$ is purely discrete, i.e., if $\rho^c\equiv 0$, then 
the set of atoms can be considered as a new set of states. This reduces 
the problem to the case of countable $\K$. In particular, the results
of previous subsection can be considered in a more general framework 
of general Polish space $\K$ and countably-supported $\rho\in\Delta(\K)$.

The following theorem describes the asymptotic behavior of the maximal variation for uncountable Polish space $\K$ (for example, $\K=\R$).
\begin{theorem}\label{th_maximal_variation_continuous}
Suppose $\K$ is a Polish space, and 
$\rho\in\Delta(\K)$. Then
\begin{equation}\label{eq_Psi_cont_asymp}
\Psi_N(\rho)=2\rho^c(\K) N+o(N),\quad N\to\infty.
\end{equation} 
\end{theorem}
Hence, if a nontrivial continuous component is presented, then the maximal variation
grows linearly with $N$. Theorem~\ref{th_maximal_variation_continuous} is proved
in the end of Section~\ref{sect_anomalous}.

One can easily show that Theorem~\ref{th_cavu} of R.~Aumann and M.~Maschler holds for any $4$-tuple from $\gammaclass(\K)$ with arbitrary Polish space $\K$. But 
for a prior distribution $\rho$ with nonzero
continuous component the statement
becomes almost meaningless because now the upper bound on the error term has the same order as the leading term.
Nevertheless, the theorem keeps giving the correct maximal order of magnitude of the error term (as it is for countable $\K$; see Theorem~\ref{th_alpha_Gamma}).
\begin{theorem}\label{th_error_term_continuous_prior}
Suppose  $\K$ is a Polish space. Then there exists a $4$-tuple $\G\in\gammaclass(\K)$ such that for any $\rho\in\Delta(\K)$ 
\begin{equation}
\label{eq_error_can_grow_linearly}
\liminf_{N\to\infty}{\err[\G_N(\rho)]}\geq\frac{1}{2}\rho^c(\K).
\end{equation}
 \end{theorem}
To show this a version of the game $\G_N(\rho)$ with $\K=[0,1]$ is described in Subsection~\ref{subsect_extension_of_G}.
This game has a pathological property. The one-stage payoff function is 
discontinuous at every point as a function of $k\in\K$. This observation suggests
that in order to avoid pathological situations for uncountable $\K$ one should consider games with some regularity of one-stage payoffs with respect to states.

\section{The maximal variation: upper bounds}\label{sect_from_above}
Here we prove Theorem~\ref{th_maximal_variation_upper_estimate} and derive an upper bound on the maximal variation for uncountable $\K$.  
We begin with some auxiliary estimates. 
\subsection{Scalar martingales}\label{subsect_scalar_martingales}
Let $X=(X_n, \F_n)_{n\geq 0}$ be a martingale taking values in $[0,\,1]$
with non-random $X_0=p$.
Denote by $\M_{[0,1]}(p)$ the set of all such martingales.
The $N$-term $\L^1$-variation of a scalar martingale $X\in\M_{[0,1]}(p)$ is given by 
$$\V_N(X)=\expectation\left(\sum_{n=0}^{N-1}|X_{n+1}-X_n|\right).$$
Asymptotic behavior of the maximal $\L^1$-variation  $\psi_N(p)=\sup_{X\in\M_{[0,1]}(p)}\V_N(X)$ as $N\to\infty$ was studied by
J.-F.~Mertens and S.~Zamir in~\cite{MZVariation}.  

They analyzed the limiting Bellman equation connecting $\psi_{N+1}$ with $\psi_N$ and obtained that
\begin{equation}\label{eq_MZ_asymptotics}
\psi_N(p)=\sqrt{N}\phi(x_p)(1+o(1)),
\end{equation}
where $\phi(x)=\frac{1}{\sqrt{2\pi}}e^{-\frac{x^2}{2}}$ is the standard normal density and $x_p$ is its $p$-quantile, i.e., $\int_{-\infty}^{x_p}\phi(x)dx=p$.
In~\cite{DeM1998}, B.~De~Meyer derived the same asymptotics from probabilistic arguments.
He introduced a new representation of the variation that reduces the initial problem
to investigation of the terminal distribution of an auxiliary martingale $S$.
The normal distribution then arises from a central limit theorem applied to $S$.

\begin{remark}
Let $\K$ be countable. Consider a martingale $\mu=(\mu_n,\F_n)_{n\geq 0}\in\M_{\Delta(\K)}(\rho)$. 
For any 
$k\in\K$ the scalar martingale $\mu(\{k\})=(\mu_n(\{k\}),\F_n)_{n\geq 0}$ is in $\M_{[0,1]}(\rho(\{k\}))$ and $\V_N(\mu)=
\sum_{k\in\K}\V_N\big(\mu(\{k\})\big).$
Therefore,
$\Psi_N(\rho)\leq\sum_{k\in\K}\psi_N\big(\rho(\{k\})\big).$
Note that the martingales $\mu(\{k\})$ should fulfill the constraint
$\sum_{k\in\K}\mu_n(\{k\})=1$ for any $n\geq 0$ almost surely, and this is why the last formula
is not equality. If $\K$ is finite, then~(\ref{eq_MZ_asymptotics}) implies that
\begin{equation}\label{eq_Psi_and_psi}
\limsup_{N\to\infty}\frac{\Psi_N(\rho)}{\sqrt{N}}\leq\sum_{k\in\K}\phi\big(x_{\rho(\{k\})}\big).
\end{equation}
Let us look at the right-hand side of (\ref{eq_Psi_and_psi}) separately of the inequality itself.
It can be easily checked that $\frac{\phi(x_p)}{\sqrt{2}p\sqrt{\ln\frac{1}{p}}}\to 1$ as $p\to+0$, and
thus for countably infinite $\K$ 
the sum converges iff $\Z_0(\rho)<\infty$. 
This observation makes the answer to Neyman's question about the class of 
$\rho$ with $\sqrt{N}$-behavior of $\Psi_N(\rho)$ rather intuitive (see Section~\ref{sect_results}).
But there is an obstacle for making this reasoning completely rigorous.
From \cite{MZVariation}
   and \cite{DeM1998}
it can be deduced only that there is an absolute constant $C>0$ such that
$$\frac{\psi_N(p)}{\sqrt{N}}\leq\phi(x_p)+C N^{-q},$$
where $q=\frac{1}{2}$ in the first paper and $q=\frac{1}{4}$ in the second.
The term $C N^{-q}$ at the right-hand side prevents direct derivation of inequality~(\ref{eq_Psi_and_psi}) in the case of infinite $\K$. This also explains why Theorem~\ref{th_maximal_variation_alpha_Psi} is not a 
direct corollary of \cite{MZVariation}
   and \cite{DeM1998}
even if $\varepsilon^*(\rho)=0$. 
\end{remark} 

To overcome the  mentioned difficulty we derive an estimate on $\V_N(X)$ without additional
term at the cost of worsening the constant.
\begin{proposition}\label{lm_scalar_bound}
For any $p\in[0,1]$ and $X\in\M_{[0,1]}(p)$ the following estimates hold:
\begin{equation}
\label{eq_variation_bound_for_scalar_theorem}
\V_N(X)\leq \sqrt{2 N}p\sqrt{\ln\frac{1}{p}}\left(1+\frac{1}{2 \ln 2}\right); 
\end{equation}
\begin{equation}
\label{eq_rough_variation_bound_for_scalar}
\V_N(X)\leq 2 N p.
\end{equation}
\end{proposition}
To infer (\ref{eq_variation_bound_for_scalar_theorem}) De~Meyer's approach is used. But instead of the central limit theorem we apply
large deviation estimates.

Let $Y_n$ be a random variable equal to $1$, when $X_{n}\geq X_{n-1}$, and $-1$, otherwise.
The auxiliary process $S$ corresponding to a martingale $X=(X_n,\F_n)_{n\geq 0}$ is defined by
$S_n=\sum_{m=1}^n Z_m$,
where $Z_m=Y_m-\expectation(Y_m\mid\F_{m-1})$.
Obviously, $S=(S_n, \F_{n})_{n\geq 0}$ is a martingale. 
In~\cite{DeM1998}, B.~De~Meyer proved that
\begin{equation}
\label{eq_DeM_scalar_representation}
\V_N(X)=\expectation X_N S_N.
\end{equation}
This identity allows to derive upper bounds on the variation from tail estimates for $S_N$.
By the standard technique one can prove the following lemma.
\begin{lemma}\label{lm_S_tails_bound}
$\forall t\geq 0\quad \P(\{S_N\geq t\})\leq\exp\left(-\frac{t^2}{2N}\right).$
\end{lemma}
Note that the result with worse constant follows immediately from 
the Bernstein or the Azuma-Hoeffding inequalities (see \cite{Azuma}).
\begin{proof}
The proof is based on the exponential Chebyshev inequality. For any $\lambda>0$ the Chebyshev inequality implies
$$\P(\{S_N\geq t\})=\P(\{\exp(\lambda S_N)\geq \exp(\lambda t)\})\leq\exp(-\lambda t)\expectation\exp(\lambda S_N).$$
From the martingale property it follows that 
$$\expectation\exp(\lambda S_n)=\expectation\big(\exp(\lambda S_{n-1})\expectation(\exp(\lambda Z_n)\mid\F_{n-1})\big).$$
One can easily prove that 
$q e^{2\lambda (1-q)}+(1-q) e^{-2\lambda q}\leq e^{\frac{\lambda^2}{2}}$
for any $q\in[0,1]$. Therefore, denoting 
$\P(\{Y_n=1\}\mid\F_{n-1})$ by $q$
we get 
$$\expectation(\exp(\lambda Z_n)\mid\F_{n-1})\leq \exp(\lambda^2/2).$$
Hence, $\expectation\exp(\lambda S_N)\leq\exp(N\lambda^2/2)$, and
choosing $\lambda=t/N$ concludes the proof. 
\end{proof}

We estimate $\expectation X_N S_N$ by maximizing
this expectation under constraints given by the information we have about the
distributions of $X_n$ and $S_n$. Note that the 
problem of maximizing $\expectation XY$ over all joint distributions with
prescribed marginals is called
the maximal covariance problem. 
Our proof of Lemma~\ref{lm_scalar_bound} is close to the proof of Theorem~6 from \cite{DeM2010}, where a solution to the maximal covariance problem is described.
\begin{proof}[Proof of Proposition~\ref{lm_scalar_bound}]
First we obtain the estimate~(\ref{eq_variation_bound_for_scalar_theorem}) for
$0<p\leq\frac{1}{2}$
(if $p=0$, the estimate holds trivially). 
Consider the auxiliary process $S$ and denote by $G$ the cumulative distribution function of $S_N/\sqrt{N}$.
For a monotonically increasing function $H$ its ``inverse'' is defined by $H_{\mathrm{inv}}(z)=\sup\{t\mid H(t)\leq z\}$.
De~Meyer's representation
(\ref{eq_DeM_scalar_representation}) implies
$$\V_N(X)=\sqrt{N}\int_{-\infty}^\infty t\expectation(X_N\mid S_N=t\sqrt{N}) dG(t)=\sqrt{N}\int_{0}^1 G_{\mathrm{inv}}(z)h(z)\,dz,$$
where $z=G(t)$ and  $h(z)=\expectation(X_N\mid S_N=\sqrt{N}G_{\mathrm{inv}}(z))$.
We set $F(t)=1-e^{-\frac{t^2}{2}}$ for $t\geq 0$ and $F(t)=0$, otherwise.
Lemma~\ref{lm_S_tails_bound} implies $\ F(t)\leq G(t)$ for all $t\in\R$, and,
hence,  $G_{\mathrm{inv}}(z)\leq F_{\mathrm{inv}}(z)$ for all
$z\in[0,1]$.
Therefore, $\V_N(X)\leq\sqrt{N}\int_{0}^1 F_{\mathrm{inv}}(z)h(z)\,dz.$
Note that $h(z)\in[0,1]$, and $\int_{0}^1 h(z)\,dz=p$ since $\expectation X_N=p$.
Denote by $W(p)$ the set of all such functions.
Hence, $\V_N(X)\leq\sqrt{N}\max_{f\in W(p)}\int_{0}^1 F_{\mathrm{inv}}(z)f(z)\,dz$.
The set $W(p)$ is convex and optimization problem is linear. Therefore, the 
maximum is attained at a peak point of $W(p)$. Peak points are
indicator functions of subsets $B\subset[0,1]$ of Lebesgue measure $p$. 
From monotonicity of $F_{\mathrm{inv}}$ we get
$$\V_N(X)\leq\sqrt{N}\int_{1-p}^1 F_{\mathrm{inv}}(z)\,dz=\sqrt{N}\int_{t_p}^\infty t\, dF(t),$$
where $t_p=\sqrt{2\ln\frac{1}{p}}$ is the solution of $1-p=F(t_p)$. Integration by parts implies
$\int_{t_p}^\infty t\, dF(t)=t_p e^{-\frac{t_p^2}{2}}+\int_{t_p}^\infty e^{-\frac{t^2}{2}}\,dt.$
The integral at the right-hand side can be estimated in the following way
$$\int_{t_p}^\infty e^{-\frac{t^2}{2}}\,dt=\int_{0}^\infty e^{-\frac{t^2+t_p^2+2 t_p t}{2}}\,dt\leq e^{-\frac{t_p^2}{2}}\int_{0}^\infty e^{t_p t}\,dt=\frac{1}{t_p}e^{-\frac{t_p^2}{2}}.$$
Taking into account that $\left({2 \ln\frac{1}{p}}\right)^{-1}\leq\left({2 \ln 2}\right)^{-1}$,  we get the estimate (\ref{eq_variation_bound_for_scalar_theorem})  for $p\leq \frac{1}{2}$.

Now we consider $p\in\left(\frac{1}{2},1\right]$. Note that
if $X\in\M_{[0,1]}(p)$, then $X'$ given by $X'_n=1-X_n$ belongs to 
$\M_{[0,1]}(1-p)$, and $\V_N(X)=V_N(X')$. Put $\varphi(p)=p\sqrt{\ln\frac{1}{p}}$. Therefore, (\ref{eq_variation_bound_for_scalar_theorem}) follows from the already considered case by the elementary
inequality $\varphi(p)\geq \varphi(1-p)$ for $p\in\left[\frac{1}{2},1\right]$.

The rough estimate (\ref{eq_rough_variation_bound_for_scalar}) immediately 
follows from $|X_{n+1}-X_n|\leq X_{n+1}+X_n$ and $\expectation X_n= p$.

\end{proof}
\subsection{$\Delta(\K)$-valued martingales: discrete $\rho$}
Suppose $\K$ is countable.
Then   
$\V_N(\mu)=
\sum_{k\in\K}\V_N\big(\mu(\{k\})\big)$
for any
$\mu\in\M_{\Delta(\K)}(\rho)$.
As we will see, the appropriate upper bound from Proposition~\ref{lm_scalar_bound} applied to scalar martingales at the right-hand side of this identity
implies Theorem~\ref{th_maximal_variation_upper_estimate}. 
So in the proof we do not use that 
$\sum_{k\in\K}\mu_n(\{k\})\equiv 1$ but only that the sum of expectations
equals $1$. One could expect that this approach gives a very
rough estimate. Rather surprisingly the resulting upper bound
reflects the correct order of magnitude of $\Psi_N(\rho)$ (see Theorem~\ref{th_maximal_variation_alpha_Psi}).
\begin{proof}[Proof of Theorem~\ref{th_maximal_variation_upper_estimate}]
Fix arbitrary $\mu\in\M_{\Delta(\K)}(\rho)$, $N\in\N$, and  $\varepsilon\in[0,1/2]$
such that $\Z_\varepsilon(\rho)$ is finite. It is enough
to show that 
$\V_N(\mu)\leq c N^{\frac{1}{2}+\varepsilon}\Z_\varepsilon(\rho)$.
We represent $K$ as the 
union of $\K_{\leq}=\left\{k\in\K\mid\ln\big(1/\rho(\{k\})\big)\leq N\right\}$ and
$\K_{>}=\K\setminus\K_{\leq}$.
Applying (\ref{eq_variation_bound_for_scalar_theorem}) to the contribution
of $k\in\K_\leq$ and taking into account the definition of $\K_\leq$  we obtain
$$\sum_{k\in\K_\leq}\V_N\big(\mu(\{k\})\big)
\leq
\sqrt{2}\left(1+\frac{1}{2\ln 2}\right)N^{\frac{1}{2}+\varepsilon}\sum_{k\in\K_{\leq}}\rho(\{k\})\left(\ln\frac{1}{\rho(\{k\})}\right)^{\frac{1}{2}-\varepsilon}.$$
For $k\in\K_>$ estimate (\ref{eq_rough_variation_bound_for_scalar}) implies
$$\sum_{k\in\K_>}\V_N\big(\mu(\{k\})\big)\leq
2 N^{\frac{1}{2}+\varepsilon}\sum_{k\in\K_{>}}\rho(\{k\})\left(\ln\frac{1}{\rho(\{k\})}\right)^{\frac{1}{2}-\varepsilon}.$$
Since $2\leq \sqrt{2}\left(1+\frac{1}{2\ln 2}\right)$,
we get
$$\V_N(\mu)=\sum_{k\in\K}\V_N\big(\mu(\{k\}\big)\leq \sqrt{2}\left(1+\frac{1}{2\ln 2}\right)N^{\frac{1}{2}+\varepsilon}\Z_\varepsilon(\rho).$$
This completes the proof.
\end{proof}
\begin{remark}\label{rem_sublinear_growth}
We claim that the growth of the maximal variation $\Psi_N(\rho)$ is sublinear even if $\Z_\varepsilon(\rho)=\infty$ for any $\varepsilon<\frac{1}{2}$. Indeed, let 
$K_\delta$ be a subset of $\K$ such that $\rho(\K_\delta)\leq\delta$.
Applying (\ref{eq_variation_bound_for_scalar_theorem}) for 
$k\in\K\setminus\K_\delta$ and (\ref{eq_rough_variation_bound_for_scalar}) for
$k\in\K_\delta$ we get
$\Psi_N(\rho)\leq C_\delta\sqrt{N}+2\delta N$. Since $\delta>0$ is arbitrary,
$$\lim_{N\to\infty} \frac{\Psi_N(\rho)}{N}=0.$$
\end{remark}
\subsection{$\Delta(\K)$-valued martingales: $\rho$ with nontrivial continuous component}
This subsection is devoted to the case of uncountable $\K$. 
\begin{proposition}\label{prop_maximal_variation_upper_estimate_continuous}
Suppose $\K$ is a Polish space,  
and $\rho\in\Delta(\K)$ is decomposed as $\rho^c+\rho^d$. Then
\begin{equation}\label{eq_Psi_cont_upper_bound}
\Psi_N(\rho)\leq 2\rho^c(\K)N+\Psi_N(\rho^{d+}),
\end{equation} 
where $\rho^{d+}=\rho^d+(1-\rho^d(\K))\delta_{k_0}$, and 
$\delta_{k_0}$ is the Dirac $\delta$-measure concentrated at a point $k_0\in\K$
such that $\rho^d(\{k_0\})=0$.
\end{proposition}
\begin{proof}
Consider an arbitrary martingale $\mu=(\mu_n,\F_n)_{n\geq 0}\in\M_{\Delta(\K)}(\rho)$. Let $\K'$ be the set of atoms
of $\rho$. Define the processes of continuous and discrete parts of $\mu$ by
$\mu_n^c(B)=\mu_n\big(B\cap(\K\setminus\K')\big)$ and 
$\mu_n^d(B)=\mu_n\big(B\cap\K'\big)$ for any Borel $B\subset\K$, respectively.
Since $\mu^c_n$ and $\mu^d_n$ are mutually singular,
$\V_N(\mu)=\V_N(\mu^c)+\V_N(\mu^d)$ (the variation can be obviously extended
to the processes with values in finite measures).
Note that $\|\Phi-\Phi'\|_\tv\leq\Phi(\K)+\Phi'(\K)$ for any finite 
positive measures $\Phi$ and $\Phi'$. Hence, by martingale property
$\V_N(\mu^c)\leq 2N\rho^c(\K)$. Define $\mu^{d+}$
by $\mu_n^{d+}=\mu_n^{d}+(1-\mu_n^{d}(\K))\delta_{k_0}$. Therefore,
$\mu^{d+}\in\M_{\Delta(\K)}(\rho^{d+})$, and $\V_N(\mu^{d})\leq\V_N(\mu^{d+})\leq\Psi_N(\rho^{d+})$. This concludes the proof.

\end{proof}
\begin{remark}\label{rem_why_discrete_part_can_be_neglected}
The measure $\rho^{d+}$ is countably supported, and,
therefore, the results of previous subsection can be applied. In particular, $\Psi_N(\rho^{d+})/N\to 0$ as $\quad N\to\infty$. Also note that if 
$\rho$ is purely continuous, then $\Psi_N(\rho^{d+})=0$.
\end{remark}
\section{The maximal variation: anomalous growth}\label{sect_anomalous}
In this section
a $\Delta(\K)$-valued martingale $\Upsilon^\rho$ with
rapidly growing variation is described. For this purpose 
we extend the construction from the paper~\cite{Neyman} of A.~Neyman.
As a corollary we get 
a lower bound on the maximal variation with correct
order of magnitude for large $N$. This lower bound together 
with the results of the previous section allow us to prove 
Theorems~\ref{th_maximal_variation_alpha_Psi}~and~\ref{th_maximal_variation_continuous}.

Consider a sequence $\{\omega_n\}_{n=1}^\infty$ of i.i.d. Bernoulli random variables
with success probability $1/2$. Let $(\F_n^2)_{n\geq 0}$ be the filtration they generate
(the so-called dyadic filtration). A martingale adapted to $(\F_n^2)_{n\geq 0}$ is also called dyadic. The set of all dyadic martingales from $\M_{\Delta(\K)}(\rho)$ is denoted by $\M_{\Delta(\K)}^2(\rho)$.
Denote by $\one_B$ the indicator function of a set $B$. 
For two finite signed measures $\Phi_{i}$, $i=1,2$, we write $\Phi_1=f\Phi_2$  if
$\frac{d\Phi_1}{d\Phi_2}=f$.
Let $u^M$ be the uniform distribution over $2^M$-element subset of $\K$.
In \cite{Neyman}, the following 
dyadic martingale $\nu^M=(\nu^M_n,\F_n^2)_{n\geq 0}\in\M_{\Delta(\K)}(u^M)$ was constructed.
It starts from $u^M$ and evolves according to 
$$\nu^M_{n}=2\big(\one_{B_n}\omega_{n}+\one_{{\K\setminus B_n}}(1-\omega_{n})\big)\nu^M_{n-1}$$ 
for any $n=1,2,...M$; here $B_n\subset \K$ is a predictable ($\F_{n-1}^2$-measurable) random subset such that $\nu^M_{n-1}(B_n)=1/2$.
For $n\geq M$ the martingale $\nu^M$ is constant, i.e, $\nu^M_n=\nu_M^M$. The main feature of $\nu^M$ is that
$\V_N(\nu^M)=N$ for $N\leq M$, i.e., the variation of $\nu^M$ grows linearly
if $N$ is not too large. 
\subsection{Martingale of dyadic splittings for $\rho\in\Delta(\R)$}
A dyadic martingale 
$\Upsilon^\rho$ described here extends the construction of $\nu^M$ to an arbitrary prior distribution
$\rho$ over the real line. 

The minimal closed
support of $\rho\in\Delta(\R)$ is the complement to the union of open sets of zero measure and is denoted by $\supp\rho$.
Consider a pair of transformations $\pi_{i}: \Delta(\R)\to\Delta(\R),\ \ i=0,1$,
such that for any $\rho\in\Delta(\R)$
\begin{equation}\label{eq_pi_properties}
\rho=\frac{1}{2}\pi_0[\rho]+\frac{1}{2}\pi_1[\rho],\quad \mbox{and} \quad \supp \pi_0[\rho]\leq\supp\pi_1[\rho]
\end{equation} 
(for $B_{1},B_{2}\subset\R$ we write $B_1\leq B_2$ if $\forall x_1\in B_1, \forall x_2\in B_2\ \  x_1\leq x_2$).

Such transformations exist, and one can show that they are uniquely determined by~(\ref{eq_pi_properties}). Let us describe
them explicitly.
A number $m_\rho=\min\left\{x\in\R\mid \rho\big((-\infty,x]\big)\geq\frac{1}{2}\right\}$
is the least median of $\rho$. 
Define a function $\lambda_\rho:\R\to[0,2]$ by the following conditions:
$\lambda_\rho(x)=2$ for $x< m_\rho $; 
$\lambda_\rho(x)=0$ for $x> m_\rho $; and
$\int_\R \lambda_\rho(x)d\rho(x)=1$.
Then we set $\pi_0[\rho]=\lambda_\rho\rho$
and $\pi_1[\rho]=2\rho-\pi_0[\rho]$. 
Roughly speaking, $\pi_{0}[\rho]$ and $\pi_{1}[\rho]$ are the normalized restrictions of $\rho$ to the left from the median and to the right, respectively.

We define the process $\Upsilon^\rho$ by
$$\Upsilon_{n}^\rho=\pi_{\omega_n}[\Upsilon_{n-1}^\rho]\quad \mbox{and}\quad
\Upsilon_{0}^\rho=\rho.$$
The properties (\ref{eq_pi_properties}) of $\pi_{i}$ immediately imply that
$\Upsilon^\rho=(\Upsilon_n^\rho,\F_n^2)_{n\geq 0}$ is a
dyadic martingale from $\M_{\Delta(\K)}(\rho)$. 
If $\rho=u^M$, then $\Upsilon^\rho$ and $\nu^M$ coincide.
\begin{remark}\label{rem_Upsilon_another_construction}
There is a more compact alternative description of $\Upsilon^\rho$, $\rho\in\Delta(\R)$. Denote by $F$ the cumulative distribution function corresponding to $\rho$. Let $z$ be a random number uniformly distributed over $[0,1]$. Hence, $k=\sup_t\{F(t)\leq z\}$ is $\rho$-distributed. Consider a binary representation
$z=0.\omega_1\omega_2\omega_3...$ Then $\{\omega_n\}_{n=1}^\infty$ generate the so-called standard dyadic filtration $(\F_n^2)_{n\geq 0}$ of $[0,1]$. Finally, we define $\Upsilon_n^\rho$ as the conditional distribution of $k$ given $\F_n^2$, i.e., $\Upsilon_n^\rho(B)=\P(\{k\in B\}\mid \F_n^2)$
for any Borel $B\subset\R$.
\end{remark}
\subsection{Countably-supported prior distributions}
\begin{proposition}\label{prop_Upsilon_variation_growth}
Suppose $\K$ is countable and $\rho\in\Delta(\K)$ is such that
$\Z_{\varepsilon}(\rho)=\infty$ for some $\varepsilon\in\left[0,\frac{1}{2}\right)$. Then 
there exists a dyadic $\Delta(\K)$-valued martingale 
$\mu\in\M_{\Delta(\K)}(\rho)$ such that
for any $\gamma<\varepsilon$
\begin{equation}\label{eq_Upsilon_variation_asymp}
\limsup_{N\to\infty}\frac{\V_N(\mu)}{N^{\frac{1}{2}+\gamma}}=\infty.
\end{equation}
\end{proposition}
Without loss of generality, $\K=\N$, and $\rho$ is monotonically
decreasing, i.e., $\rho(\{k\})$ is greater than $\rho(\{k+1\})$ for all $k\in\N$ (indeed, starting from general $\K$ we can enumerate its elements in a suitable way). We identify probabilities on $\N$ and probabilities on $\R$ supported on $\N$. The objective is to show that in this case we can take
$\mu=\Upsilon^\rho$. Several lemmas precede the proof.

Denote $\V_N(\Upsilon^\rho)$ by $Q_N(\rho)$. We need a lower estimate on $Q_N(\rho)$,
but the situation differs from considered in A.~Neyman's paper~\cite{Neyman}. 
For the martingale $\nu^M$ the two possible values of $\nu^M_{n+1}$ given $\nu^M_n$ are mutually singular. This is why
each stage $n\leq M$ contributes $1$ to the variation of $\nu^M$. 
Bot not every probability distribution can be represented as the arithmetic mean of two mutually
singular probability distributions (for example, Bernoulli distribution with success probability
different from $1/2$). In particular, the two values
of $\Upsilon_{n+1}^\rho$ given $\Upsilon_n^\rho$ ($\pi_{0}[\Upsilon_n]$ and $\pi_{1}[\Upsilon_n]$) 
are not mutually singular in general case, and this makes the problem more complicated.
We overcome this difficulty by observing that they are ``almost mutually singular'' if
all the atoms of $\Upsilon_n^\rho$ are small enough. Therefore, we should control 
the heaviest atom of $\Upsilon_n^\rho$. This is why
the condition $\Z_\varepsilon(\rho)=\infty$ arises for the lower bound on $Q_N(\rho)$ to be of the order of $N^{\frac{1}{2}+\varepsilon}$.
We set $H(\rho)=\max_k \rho(\{k\})$.
\begin{lemma}\label{lm_Upsilon_one_stage_variation}
$1-H(\rho)\leq Q_1(\rho)\leq 1$ for any $\rho\in\Delta(\R)$.
\end{lemma} 
\begin{proof}
By the definition $Q_1(\rho)=\frac{1}{2}\|\pi_0[\rho]-\rho\|_\tv+\frac{1}{2}\|\pi_1[\rho]-\rho\|_\tv$. Using the property (\ref{eq_pi_properties}) we get
$Q_1(\rho)=\frac{1}{2}\|\pi_0[\rho]-\pi_1[\rho]\|_\tv$. Note that $\pi_0[\rho]$ and
$\pi_1[\rho]$ can have a joint atom at $k=m_\rho$. ``The worst case'' is
$\pi_0[\rho](\{m_\rho\})=\pi_1[\rho](\{m_\rho\})$. This implies the desired lower
bound $Q_1(\rho)\geq 1-\rho(\{m_\rho\})\geq 1-H(\rho)$. The upper bound corresponds to ``the best case'' of mutually singular $\pi_0[\rho]$ and
$\pi_1[\rho]$.
\end{proof}
From the recurrent structure of $\Upsilon^\rho$ we immediately get the following identity.
\begin{lemma}\label{lm_Upsilon_identity_for_variation}
$Q_{M+N}(\rho)=Q_{M}(\rho)+\expectation Q_N(\Upsilon_{M}^\rho)$ for any $\rho\in \Delta(\R)$ and $M,N\in\N$.
\end{lemma} 
The next estimate follows from Lemmas~\ref{lm_Upsilon_one_stage_variation}~and~\ref{lm_Upsilon_identity_for_variation}.
\begin{lemma}\label{lm_Upsilon_variation_from_below}
$N-\expectation\left(\sum_{n=0}^{N-1}H(\Upsilon_n^\rho)\right)\leq Q_N(\rho)\leq N.$
\end{lemma}
For any $d=(d_1,d_2,...,d_M)\in\{0,1\}^M$ we denote the composition $\pi_{d_M}\circ\pi_{d_{M-1}}\circ...\circ\pi_{d_1}$  by $\pi(d)$. Let $N(\rho,d)$ be the integer part of
$-\log_2H(\pi(d)[\rho])$.
\begin{lemma}\label{lm_Upsilon_variation_from_below_extended}
$Q_{M+N(\rho,d)}(\rho)\geq {2^{-M}}(N(\rho,d)-1)\quad \forall M\in \N,\ d\in\{0,1\}^M$.
\end{lemma}
\begin{proof}
Since $H(\Upsilon_n^\rho)\leq 2^n H(\rho)$, from Lemma~\ref{lm_Upsilon_variation_from_below} it follows that $Q_N(\rho)\geq N-H(\rho) 2^{N}$.
By Lemma~\ref{lm_Upsilon_identity_for_variation} we get
$Q_{M+N}(\rho)\geq\P(\{\Upsilon_M^\rho=\pi(d)[\rho]\})Q_N(\pi(d)[\rho])$, and the definition
of $\Upsilon^\rho$ implies $\P(\{\Upsilon_M^\rho=\pi(d)[\rho]\})={2^{-M}}$.

\end{proof}
Proposition~\ref{prop_Upsilon_variation_growth} is proved by selecting an
appropriate sequence $d^{(M)}\in\{0,1\}^M$, $M\in\N$, and then applying
Lemma~\ref{lm_Upsilon_variation_from_below_extended}.
\begin{proof}[Proof of Proposition~\ref{prop_Upsilon_variation_growth}]
As it is mentioned above, we can assume that $\K=\N$ and that $\rho\in\Delta(\N)\subset\Delta(\R)$ is monotonically decreasing. We set $d^{(M)}=(1,1,...1,0)$ and denote $N(\rho,d^{(M)})$ by $N_M$. By Lemma~\ref{lm_Upsilon_variation_from_below_extended} it is enough to show
that $$\limsup_{M\to\infty}\frac{2^{-M}N_M}{(N_M+M)^{\frac{1}{2}+\gamma}}=\infty.$$
Moreover, it is sufficient to check only that 
\begin{equation}\label{eq_Upsilon_condition_to_check}
\limsup_{M\to\infty}2^{-M}\left(N_M\right)^{\frac{1}{2}-\gamma}=\infty.
\end{equation}
 Indeed, for any
 sequence $M_L\to\infty$ such that $2^{-M_L}\left(N_{M_L}\right)^{\frac{1}{2}-\gamma}\to\infty$
 we also have $M_L+N_{M_L}=N_{M_L}(1+o(1))$.
 We put $\rho_M=\pi(d^{(M)})[\rho]=\pi_0(\pi_1)^{M-1}[\rho]$. From the monotonicity
 of $\rho$ we get $\rho(\{k\})\geq 2^{-M-1}H(\rho_{M+1})$ for any $k\in\supp\rho_M$.
Since  $\rho$ can be represented as $ \sum_{M=1}^\infty 2^{-M}\rho_M$,  
the condition $\Z_\varepsilon(\rho)=\infty$ implies the
divergence of the sum $$\sum_{M=1}^\infty 2^{-M}\left(\ln \frac{2^{M+1}}{H(\rho_{M+1})}\right)^{\frac{1}{2}-\varepsilon}.$$ 
As
$(\ln ab)^c\leq (2 \ln b)^c+(2\ln a)^c$ for positive $c$ and $a,b\geq 1$,
we see that the sum   
$\sum_{M=1}^\infty 2^{-M}\left(N_M\right)^{\frac{1}{2}-\varepsilon}$ also diverges. This implies
(\ref{eq_Upsilon_condition_to_check}) and concludes the proof.
\end{proof}
Define
the maximal variation of dyadic martingales by
$\Psi_N^2(\rho)=\sup_{\mu\in\M_{\Delta(\K)}^2}\V_N(\mu)$. 
\begin{proposition}\label{prop_dyadic_martingales_maximal_variation}
Suppose $\K$ is countable.
Then for any non-degenerate $\rho\in\Delta(\K)$ 
and $\gamma<\varepsilon^{*}(\rho)$ we have
\begin{equation}\label{eq_maximal_variation_dyadic_asymp}
\limsup_{N\to\infty}\frac{\Psi_N^2(\rho)}{N^{\frac{1}{2}+\gamma}}=\infty.
\end{equation}
\end{proposition}
\begin{proof}
If $\varepsilon^{*}(\rho)>0$, then
$\Z_\varepsilon(\rho)=\infty$ for any $\varepsilon$
between $\gamma$ and $\varepsilon^{*}(\rho)$, and, therefore,
the statement follows from Proposition~\ref{prop_Upsilon_variation_growth}.
Now assume that
$\varepsilon^{*}(\rho)=0$. It is enough to show that there is a sequence of 
martingales $\mu^{(N)}$ from $\M_{\Delta(\K)}^2(\rho)$ such that
$\limsup_{N\to\infty} \V_N(\mu^{(N)})/\sqrt{N}>0$. 
Let $k^*$ be the heaviest atom of $\rho$. Denote $\rho(\{k^{*}\})$ by $p$.
It follows from the results of B.~De~Meyer (see~\cite{DeM1998},
the end of Section~1) that for any $p\in(0,1)$ there is a sequence of 
scalar dyadic martingales $X^{(N)}\in\M_{[0,1]}(p)$ such that
$\V_N(X^{(N)})/\sqrt{N}\to\phi(x_p)>0$, where $\phi(x_p)$ is the normal density
at its $p$-quantile. 
Finally we set $\mu_n^{(N)}=\big(\one_{\{k^*\}}X_n^{(N)}+\one_{\K\setminus\{k^*\}}(1-X_n^{(N)})\big)\rho$. Thus $\V_N(\mu^{(N)})/\sqrt{N}\to 2\phi(x_p)>0$.
\end{proof}
Now we pass to the proof of Theorem~\ref{th_maximal_variation_alpha_Psi}.
\begin{proof}[Proof of Theorem~\ref{th_maximal_variation_alpha_Psi}]
From Theorem~\ref{th_maximal_variation_upper_estimate}
it follows that
$\alpha_\Psi(\rho)\leq \frac{1}{2}+\varepsilon^*(\rho)$.
Since $\Psi_N(\rho)\geq \Psi_N^2(\rho)$,
Proposition~\ref{prop_dyadic_martingales_maximal_variation}
implies the reverse inequality.
\end{proof}
\subsection{Prior distributions with nontrivial continuous component}
In this subsection Theorem~\ref{th_maximal_variation_continuous}
 is proved. We begin with a lemma describing the case with no atoms.
\begin{lemma}\label{lm_purely_cont_prior}
Let $\rho\in\Delta(\R)$ be purely continuous. Then $\Psi_N(\rho)=2N$.
\end{lemma} 
\begin{proof}
By Proposition~\ref{prop_maximal_variation_upper_estimate_continuous}
it is enough to show that $\Psi_N(\rho)\geq 2N$.
For $D\in\N$ define a martingale
$\Upsilon^{\rho,D}$ by $\Upsilon^{\rho,D}_n=\Upsilon^{\rho}_{D n}$.
For this martingale all $2^D$ possible values of 
$\Upsilon^{\rho,D}_{n+1}$ given $\Upsilon^{\rho,D}_{n}$ are mutually singular.
The same reasoning as in Lemma~\ref{lm_Upsilon_one_stage_variation}
leads to $\V_N(\Upsilon^{\rho,D})=2N(1-2^{-D})$. Therefore,
$\Psi_N(\rho)\geq 2N(1-2^{-D})$. Since $D$ is arbitrary,
the proof is completed.
\end{proof}
\begin{proof}[Proof of Theorem~\ref{th_maximal_variation_continuous}]
The maximal variation does not depend on geometry of $\K$, and so only
the measurable structure is important. Since all uncountable Polish spaces are Borel isomorphic,  it is enough to consider a particular one, and we choose $\K=\R$. 

Let us show that
$\Psi_N(\rho)\geq 2N\rho^c(\R)$ for $\rho\in\Delta(\R)$.
If $\rho$ is discrete, this inequality is trivial. Otherwise
we set $\rho'=\frac{1}{\rho^c(\R)}\rho^c\in\Delta(\R)$. Note that for any martingale
$\mu'\in\M_{\Delta(\R)}(\rho')$ the martingale
$\mu$ given by $\mu_n=\rho^c(\R)\mu_n'+\rho^d$ belongs to $\M_{\Delta(\R)}(\rho)$, and
$\V_N(\mu)=\rho^c(\R)\V_N(\mu')$. Therefore, $\Psi_N(\rho)\geq\rho^c(\R)\Psi_N(\rho')$,
and the required inequality follows from Lemma~\ref{lm_purely_cont_prior}.
Proposition~\ref{prop_maximal_variation_upper_estimate_continuous} gives the upper 
bound. Together with Remark~\ref{rem_why_discrete_part_can_be_neglected} 
this concludes the proof.
\end{proof}
\begin{remark}\label{rem_maximal_variation_dyadic_continuous}
Consider the dyadic case. If $\rho=\rho^c$, then $H(\Upsilon_n^\rho)=0$, and Lemma~\ref{lm_Upsilon_variation_from_below} 
implies
$\Psi_N^2(\rho)\geq N$ (in fact, one can show that in this case $\Psi_N^2(\rho)=N$). Hence, by the same reasoning as in the proof
of Theorem~\ref{th_maximal_variation_continuous} we get
$\Psi_N^2(\rho)\geq N\rho^c(\K)$ for arbitrary $\rho\in\Delta(\K)$.
\end{remark}


\section{Anomalous behavior of error term}\label{sect_game_with_anomalous}
In this section we prove Theorem~\ref{th_alpha_Gamma}
that describes the slowest possible rate of error term 
decreasing. 
 
Recall that $\Psi_N^2(\rho)$ is the maximal variation of dyadic martingales
(see Section~\ref{sect_anomalous}).
\begin{proposition}\label{prop_lower_estimate_for_error_term}
If  $\K$ is countable, then there exists a $4$-tuple $\G$ from $\gammaclass(\K)$ such that 
\begin{equation}\label{eq_error_term_from_below}
\err[\G_N(\rho)]\geq\frac{1}{2 N}\Psi_N^2(\rho)\quad \mbox{for any}\ \rho\in\Delta(\K)\ \mbox{and} \ N\in\N.
\end{equation}
\end{proposition}
Theorem~\ref{th_alpha_Gamma} immediately follows from this proposition.
\begin{proof}[Proof of Theorem~\ref{th_alpha_Gamma}]
From Theorem~\ref{th_cavu} we get
$\alpha_\gammaclass(\rho)\leq \alpha_\Psi(\rho)-1$, and  Theorem~\ref{th_maximal_variation_alpha_Psi} implies
$\alpha_\Psi(\rho)=\frac{1}{2}+\varepsilon^*(\rho)$.
From Proposition~\ref{prop_lower_estimate_for_error_term} 
it follows that
$$\alpha_\gammaclass(\rho)\geq-1+ \limsup_{N\to\infty}\frac{\ln\Psi_N^2(\rho)}{\ln N},$$
and by Proposition~\ref{prop_dyadic_martingales_maximal_variation}
this limit is greater than $\frac{1}{2}+\varepsilon^*(\rho)$.
\end{proof}
To prove Proposition~\ref{prop_lower_estimate_for_error_term} we construct
the repeated game $\G_N(\rho)$ explicitly. The proof is divided into several lemmas,
where we check the properties of $\G_N(\rho)$. 
\subsection{The game $ \G_N(\rho)$ and its properties}\label{subsect_construction_of_G}
The construction is based on two ingredients. One of them is
the well-known repeated game $G_N(p)$ with incomplete information introduced by S.~Zamir in~\cite{Zamir_sqrt_is_precise}. 
This game has two-element set of states  $\{0,1\}$ and stage payoffs given by the following matrices (Player~1 is the row-chooser)
$$a^1=\left(\begin{array}{cc}3 & -1\\ -3 & 1\end{array}\right)\ \  \mbox{ and } \ \ a^0=\left(\begin{array}{cc}2 & -2\\ -2 & 2\end{array}\right).$$
As it was shown by J.-F.~Mertens~and~S.~Zamir in~\cite{MZNormalGames},
$\err[G_N(p)]=\frac{1}{2N}\Psi_N(p)(1+o(1))$ as $N\to\infty$
for any $p\in\Delta(\{0,1\})$. 
Hence, this game exhibits $1/\sqrt{N}$-behavior of the error term, i.e., the slowest possible rate of decreasing over $\gammaclass(\{0,1\})$. 
Another ingredient is a special structure that
\begin{enumerate}
\item allows informed player to choose what (state-dependent) game he wants to play at a current stage from a sufficiently large amount of alternatives,
\item informs Player~2 of this choice.
\end{enumerate}
Importance of this structure
comes from the intuition that, if Player~1 has enough choices of different stage games, then at a current stage he can emphasize any particular part of uninformed player's lack of knowledge 
 and uniformly benefit from it. Building of games with such a structure from ``elementary'' blocks was considered by A.~Neyman in~\cite{Neyman}. 

We start with informal description of the game $\G_N(\rho)$ corresponding to the $4$-tuple $\G$.
As usual, at the beginning of the game Nature chooses a $\rho$-distributed random element $k$ from the set of states $\K$. Player~1 is informed of $k$
 and Player~2 is not.
It is convenient to represent each stage $n=1,2,...N$ of the game $ \G_N(\rho)$  as being played in two steps:
\begin{enumerate}
\item at the first step Player~1 selects a finite subset $X_n\subset\K$, and Player~2 selects a finite subset
$Y_n\subset\K$, then $X_n\cap Y_n$ is told Player~2;
\item
at the second step they play the matrix game $a^1$ if $k\in X_n$, and they
play $a^0$, otherwise.
\end{enumerate}
That is, at the second step Player~1 selects a row $x_n\in\{1,2\}$ using information he has,
 and Player~2 selects a column $y_n\in\{1,2\}$.
Before the next stage the selected actions $i_n=(X_n, x_n)$ and $j_n=(Y_n,y_n)$ are announced. The total gain of Player~1 is the expected arithmetic mean of per-step gains.
\begin{remark}\label{rem_why_do_we_need_Y}
It is reasonable for Player~2 to select $Y_n$ as big as possible. But we 
forbid him to choose infinite subsets in order to truncate his
set of actions $\J$ and, therefore, to make it countable for countably infinite $\K$. 
\end{remark}
Now we define $\G$ formally. The set of states is $\K$. The action set $\I$ of Player~1 consists of pairs
$(X,x)$, where $X$ is a finite subset of $\K$, and $x$ is an element of $\{1,2\}$. The set of actions $\J$ of Player~2 consists of pairs $(Y,y)$,
where $Y$ is a finite subset of $\K$, and $y$ is a function from $2^Y$ to $\{1,2\}$. The one-stage payoff function $A$ is given by
$\A^k_{(X,x),(Y,y)}=a_{x,y(X\cap Y)}^{\one_X(k)}.$

First we prove a version of Proposition~\ref{prop_lower_estimate_for_error_term}
for finite $\K$. For this purpose we pass from games to the martingale optimization
problem similar to the maximal variation one. 
This reduction was introduced by B.~De~Meyer in~\cite{DeM2010} to analyze a market model based on repeated games with incomplete information. His approach was extended to
general repeated games with incomplete information having finite $\K$ and Polish $\I$ and $\J$ by F.~Gensbittel in~\cite{CavU}.
\begin{theorem}[F. Gensbittel]\label{th_Genbittel}
Consider a $4$-tuple $\Gamma=({\K},\I,\J,\A)$
with finite $\K$, Polish $\I$ and $\J$, and bounded $\A$.
Define an auxiliary $4$-tuple $\Gamma^\Delta$ with the set of states $\K^\Delta=\Delta(\K)$,
action sets $\I$ and $\J$, and one-stage payoff function given by the formula
$A_{i,j}^\rho=\sum_{k\in\K}A_{i,j}^k \rho(\{k\})$ for any $\rho\in\K^\Delta$.
Then for any $\rho\in\Delta(\K)$ and $N\in\N$
\begin{equation}
\label{eq_De_Meyer_optimization_problem}
{\underline{\val}}[\Gamma_N(\rho)]=\sup_{(\mu_n,\F_n)_{n\geq 0}\in\M_{\Delta(\K)}(\rho)}\expectation\left(\frac{1}{N}\sum_{n=1}^{N}
\underline{\val}\big[\Gamma_1^\Delta(\Lambda_n)\big]\right),
\end{equation}
where 
$\Lambda_n\in\Delta(\K^\Delta)$ is the conditional 
distribution of $\mu_{n}$ given $\F_{n-1}$.
\end{theorem}
\begin{remark}\label{rem_interpretation_of_partial_inform}
The game $\Gamma^{\Delta}_N(\Lambda)$ with $\Lambda\in\Delta(\K^{\Delta})$
can be interpreted as a partial information extension of $\Gamma_N(\rho)$, i.e.,
Player~1 does not know the state $k$ exactly but receives a noisy signal $M$ with
partial information about it. So $\Lambda$ represents the distribution of informed player's beliefs after observing $M$. Note that the non-revealing game 
$\Gamma_1^\nr(\rho)$ can be clearly identified with
$\Gamma_1^\Delta(\delta_\rho)$, where $\delta_\rho$ is the Dirac $\delta$-measure
 at $\rho$. 
\end{remark}
Consider the partial-information extension
of $G_1(p)$. So $p=\big(p(\{0\}),\,p(\{1\})\big)$ is distributed according to $\lambda\in\Delta(\Delta(\{0,1\}))$.
Let $\varkappa$ be the distribution of $p(\{1\})$.
Then $G_1^\Delta(\lambda)$ is equivalent to the following game.
Nature picks a random $\varkappa$-distributed state $s=p(\{1\})\in[0,1]$ and tells it to Player~1.
Further Player~1 and Player~2 play the following matrix game
$$a^{s}=s a^1+(1-s)a^0=\left(\begin{array}{cc}s+2 & s-2 \\ -(s+2) & -(s-2)\end{array}\right).$$ 
\begin{lemma}
\label{lm_MZ_game_value}
 $\val[G_1^\Delta(\lambda)]=\min_{m\in[0,1]}\expectation|s-m|.$
\end{lemma} 
\begin{proof}
If $s$ takes only finite number of different values, then
the game $G_1^\Delta(\lambda)$ has a value by the minimax theorem.
In general case the existence of the value can be checked by standard
approximation arguments. 
Let us compute $\val[G_1^\Delta(\lambda)]$. Assume that Player~2 uses a mixed strategy $(t,1-t)$, i.e., selects
the left column with probability $t$ and the right one with probability $1-t$.
Then the optimal reply of Player~1 is to select the row
that gives him a positive gain (depending on $s$). Hence, his average gain is 
$\expectation|s+(4t-2)|$. Therefore,
$\val[G_1^\Delta(\lambda)]=\min_{t\in[0,1]}\expectation|s+(4t-2)|=\min_{m\in[0,1]}\expectation|s-m|.$
\end{proof}
\begin{remark} \label{rem_MZ_game_with_dyadic_prior} If $p$ takes only two different values $p^0$ and $p^1$
equally likely, i.e.,
 $\lambda=\frac{1}{2}\delta_{(p^0(\{0\}),p^0(\{1\}))}+\frac{1}{2}\delta_{(p^1(\{0\}),p^1(\{1\}))}$, 
 then
\begin{equation}\label{eq_gamma_value_twopoint}
\val[G_1^\Delta(\lambda)]=\frac{1}{2}\expectation \|p-\expectation p\|_\tv.
\end{equation}
\end{remark}
\begin{lemma}\label{lm_lower_estimate_for_error_term_finite_case} 
The statement of Proposition~\ref{prop_lower_estimate_for_error_term}
holds for finite $\K$.
\end{lemma}
\begin{proof}
We claim that the constructed $4$-tuple $\G$ fits the requirements of  Proposition~\ref{prop_lower_estimate_for_error_term}.
Indeed, since $\K$ is finite, the action sets $\I$ and $\J$ are also finite, and so
the games $\G_N(\rho)$ and $\G_1^\nr(\rho)$ have values by the minimax theorem. Therefore, the $4$-tuple $\G$ belongs to $\gammaclass(\K)$. 

From Lemma~\ref{lm_MZ_game_value} and Remark~\ref{rem_interpretation_of_partial_inform}
it follows that $u_G(p)=0$ for any $p$. Let us check that $u_{\G}$
is also identically zero, too. If at the first step of $\G_1^\nr(\rho)$ Player~1 selects an arbitrary
$X_1\subset\K$, then at the second step both players
know only that the probability to play $a^1$ is $\rho(X_1)$ (note that Player~2 can take $Y_1=\K$ because $\K$ is finite). Therefore,
the game at the second step is in fact $G_1^\nr(p)$ with $p=(\rho(\K\setminus X_1),\, \rho(X_1))$. Playing optimally in $G_1^\nr(p)$ both players guarantee to get at least $0$.
Therefore, $u_{\G}(\rho)=0$ and $\val[\G_N(\rho)]=\err[\Gamma_N(\rho)]$ for any $\rho$.

From Theorem~\ref{th_Genbittel} it follows that the estimate (\ref{eq_error_term_from_below})
can be proved by checking that for any $\mu\in\M_{\Delta(\K)}^2(\rho)$
$$\val[\G_1^\Delta(\Lambda_1)]\geq\frac{1}{2}\expectation\|\mu_1-\expectation\mu_1\|_\tv,$$
where $\Lambda_1$ is the distribution of $\mu_1$.
Recall that for any dyadic martingale $\mu$ the random measure $\mu_{1}$ takes only two different values, say, $\rho^0$ and $\rho^1$ equally likely.
Consider the following strategy of Player~1 in $\G_1^\Delta(\Lambda_1)$.
At the first step he selects $X_1=\K_<$, where $\K_<=\{k\in\K\mid \rho^0(\{k\})<\rho^1(\{k\})\}$.
Then at the second step the payoff
matrix is $a^s$, where $s$ equals $\rho^0(\K_<)$ or $\rho^1(\K_<)$ with probability $\frac{1}{2}$. Player~2 does not know $s$, but Player~1 does. In other words, at the second step
players face the game $G_1^\Delta(\lambda)$ with $\lambda=\frac{1}{2}\delta_{p^0}+
\frac{1}{2}\delta_{p^1}$, where $p^{i}=(\rho^{i}(\K\setminus\K_<),\,
\rho^{i}(\K_<))$, $i=0,\,1$. By (\ref{eq_gamma_value_twopoint}) the optimal behavior 
at the second step gives Player~1 at least
$\frac{1}{2}\expectation\|p-\expectation p\|_\tv=\frac{1}{2}\expectation\|\mu_1-\expectation \mu_1\|_\tv$.
This concludes the proof.
\end{proof}
To ensure that the result remains valid also for countably infinite $\K$ we use
finite approximations. For this purpose we should check that 
$\Psi_N^2(\rho)$ depends on $\rho$ in a regular way.
\begin{lemma}\label{lm_maximal_variation_is_lipschitz} If $\K$ is countable, then $\Psi_N^2(\rho)$ is a $2N$-Lipschitz function of $\rho$ in the total variation norm, i.e., 
$$|\Psi_N^2(\rho)-\Psi_N^2(\rho')|\leq 2N\|\rho-\rho'\|_\tv\quad \forall\rho,\rho'\in\Delta(\K).$$
\end{lemma}
\begin{remark}\label{rem_not_only_dyadic_variation_is_lipschitz}
The same proof shows that $\Psi_N(\rho)$ is also $2N$-Lipschitz.
\end{remark}
\begin{proof}
It is enough to show that for any $\rho\ne\rho'$ and any martingale $\mu\in\M_{\Delta(\K)}^2(\rho)$
there exists $\mu'\in\M_{\Delta(\K)}^2(\rho')$ such that
$|\V_N(\mu)-\V_N(\mu')|\leq 2N\|\rho-\rho'\|_\tv$. 
Let $\theta_n$ be a positive measure on $\K_<=\{k\in\K\mid \rho'(\{k\})<\rho(\{k\})\}$ such that  $\theta_n(\{k\})=\frac{\rho(\{k\})-\rho'(\{k\})}{\rho(\{k\})}\mu_n(\{k\})$
for $k\in\K_<$.
Define $\mu'$ on one-element sets by
$$\mu_n'(\{k\})=\mu_n(\{k\})+\left\{\begin{array}{cc}-\theta_n(\{k\}), &\ \ k\in\K_<\\
\frac{\rho'(\{k\})-\rho(\{k\})}{\rho(\K_<)-\rho'(\K_<)}\theta_n(\K_<), &\ \ k\in\K\setminus\K_<
 \end{array}\right..$$
Then it is easy to check that
$\mu'\in\M_{\Delta(\K)(\rho')}$, and $\expectation\|\mu_n-\mu_n'\|_\tv=
\|\rho-\rho'\|_\tv$ for any $n$. Thus the result follows from the triangle inequality.
\end{proof}
The following lemma is well known  (see~\cite{Bigbook}, p.219
and p.222) in the case of finite $\K,\I$ and $\J$.  
\begin{lemma}\label{lm_values_are_lipshitz}
Suppose that in a $4$-tuple $\Gamma=(\K,\I,\J,\A)$ the sets 
$\K$, $\I$, and $\J$ are arbitrary Polish spaces and
$\A$ is bounded. Then $\underline\val[\Gamma_1^\nr(\rho)]$,
$\overline\val[\Gamma_1^\nr(\rho)]$, $\underline\val[\Gamma_N(\rho)]$, and 
$\overline\val[\Gamma_N(\rho)]$ are $\|\A\|_\infty$-Lipschitz functions of $\rho$
in the total variation norm.
\end{lemma}
\begin{proof}
The idea of the proof comes from the paper \cite{CavU} of F.~Gensbittel (Proposition~2.1). 
It is enough to show that 
$|g_N(\rho,\sigma,\tau)-g_N(\rho',\sigma,\tau)|\leq \|\A\|_\infty\|\rho-\rho'\|_\tv$
for any behavioral strategies $\sigma$ and $\tau$ and for any $\rho,\rho'\in\Delta(\K)$ (recall that $g_N(\rho,\sigma,\tau)$ is defined by~(\ref{eq_g_N_definition})). 
The quantity $q_N(k_0,\sigma,\tau)=\frac{1}{N}\expectation_{\rho,\sigma,\tau}(\sum_{n=1}^N \A_{i_n,j_n}^k \mid k=k_0)$ can be regarded as unconditional expectation with respect to the probability measure over $\{k_0\}\times(\I\times\J)^N$ generated by $\sigma$ and $\tau$, and, hence, $q$ does not depend on the prior distribution $\rho$.
Therefore,
$$g_N(\rho,\sigma,\tau)-g_N(\rho',\sigma,\tau)=\int_\K(d\rho(k)-d\rho'(k))q_N(k,\sigma,\tau).$$ 
Since $q$ is bounded by $\|\A\|_\infty$ in absolute value, the proof is completed.
\end{proof}
Now we turn to the proof of Proposition~\ref{prop_lower_estimate_for_error_term}.
\begin{proof}[Proof of Proposition~\ref{prop_lower_estimate_for_error_term}]
The case of finite $\K$ is considered in Lemma~\ref{lm_lower_estimate_for_error_term_finite_case}, and, hence, we assume that
$\K$ is countably infinite and analyze $\G$ in this case.
Note that $\I$ and $\J$ are also countably infinite for such $\K$. 

We claim that for any  $\rho\in\Delta(\K)$ the games $\G_1^\nr(\rho)$ and $\G_N(\rho)$ have values and
that (\ref{eq_error_term_from_below}) holds. 
Indeed, consider a sequence $\rho_n$ of finitely-supported probabilities converging to $\rho$ in the total variation norm.
By Lemmas~\ref{lm_values_are_lipshitz}~and~\ref{lm_maximal_variation_is_lipschitz}, if we already know the result for all $\rho_n$, then 
we also get it for $\rho$. 

Therefore, the problem is reduced to the case of 
$\rho$ supported on a finite subset of $\K$, say $\K'$. We write $^\K\G_N(\rho)$
to indicate the set of states explicitly. Then there are two different games 
$^\K\G_N(\rho)$ and $^{\K'}\G_N(\rho)$. In contrast to the second game, the fist one has infinite sets of actions. In $^\K\G_N(\rho)$ both players know that $k$ belongs
to $\K'$ almost surely. Then selecting $X_n\cap \K'$ instead of $X_n$ leads to
the same game at the second step, and also choosing $Y_n\cap \K'$ instead of $Y_n$
does not affect the information Player~2 receives. Hence, if players are restricted
to use in $^\K\G_N(\rho)$ only the strategies inherited from $^{\K'}\G_N(\rho)$, 
then they get the same guarantied payoffs as without any restrictions.
Therefore, by Lemma~\ref{lm_lower_estimate_for_error_term_finite_case} $\underline\val[^\K\G_N(\rho)]=\overline\val[^\K\G_N(\rho)]=
\val[^{\K'}\G_N(\rho)]\geq\frac{1}{2N}\Psi_N^2(\rho)$ and $\underline\val[^\K\G_1^\nr(\rho)]=\overline\val[^\K\G_1^\nr(\rho)]=
\val[^{\K'}\G_1^\nr(\rho)]=0$, which concludes the proof.
\end{proof}
\subsection{Extension to uncountable $\K$}\label{subsect_extension_of_G}
Here we briefly describe a version of the game $\G_N(\rho)$ for $\K=[0,1]$.
Let $(\F_n^2)_{n\geq 0}$ be the standard dyadic filtration of $[0,1]$ with the Lebesgue measure (see Remark~\ref{rem_Upsilon_another_construction}). Denote by $\mathcal{X}$ the union
of $\F_{n}^2$ over all $n$. Each stage $n$ of $\G_N(\rho)$ consists of two steps
as it was for countable $\K$. But now at the first step
Player~1 selects $X_n\in\mathcal{X}$, and Player~2 selects $M_n\in\N$. Before the second step 
that proceeds as before the smallest set $Y_n\in\F_{M_n}^2$ containing $X_n$ is told to Player~2.
\begin{proof}[Proof of Theorem~\ref{th_error_term_continuous_prior}]
The scheme of the proof is the following. 
Analyzing the properties of $\G$ one can show that
Proposition~\ref{prop_lower_estimate_for_error_term} remains valid 
for $\K=[0,1]$ (or even for any uncountable Polish space). 
Then Remark~\ref{rem_maximal_variation_dyadic_continuous} implies  the result.
\end{proof}
\begin{remark}\label{rem_about_pathology}
The game constructed has a pathological property:
for any $k, k'\in[0,1]$ such that $k\ne k'$
we have $\sup_{i,j}|A_{i,j}^k-A_{i,j}^{k'}|=1$. In other words, the payoff
function is discontinuous at every point of $[0,1]$. This suggests that the non-decreasing error term is a kind of pathology.
\end{remark}

\section{Concluding remarks}\label{sect_conclusion}
A well-known problem in the theory of repeated games with incomplete information
is to show that after proper normalization the error term
and the maximal variation converge as $N\to\infty$. Even for finite $\K$ this problem is not solved in full generality (the main existing results can be found in \cite{DeM1996_1,DeM1996_2,DeM1998, DeM2010, Gensbittel2013,MZNormalGames, MZVariation, MZ1995}). 

Of course, this convergence problem has a counterpart concerning anomalous behavior for heavy-tailed
prior distributions $\rho$ over countable $\K$. Even for logarithmic asymptotics we 
do not know the existence of the limits. For example, can one put $\liminf$ instead of $\limsup$
in the definitions of $\alpha_\Psi(\rho)$ and $\alpha_\gammaclass(\rho)$ without changing
the statements of Theorems~\ref{th_maximal_variation_alpha_Psi}~and~\ref{th_alpha_Gamma}?

We guess that the answer to this question is ``yes''. Hence,  
$\Psi_N(\rho)=N^{\alpha_\Psi(\rho)+o(1)}$ as $N\to\infty$ for non-degenerate $\rho$.
This leads to a more delicate question about convergence of
$\Psi_N(\rho)/N^{\alpha_\Psi(\rho)}$ for large $N$. 
We expect that this limit may fail to exist for general $\rho$ such that $\alpha_\Psi(\rho)>\frac{1}{2}$. 
The reason is that the exponent $\alpha_\Psi$ is not constant 
when $\rho$ ranges over the class of such heavy-tailed distributions, and this 
suggests the possibility of ``intermediate'' behaviors (for example, $N^{o(1)}$ can contain logarithmic
growth). Our conjecture is that the proper normalization of $\Psi_N(\rho)$ should be given in terms of the ``distribution function'' $F(n)=\max_{B\subset K,\, \#B\leq n} \rho(B)$, and, hence, should precisely reflect tail asymptotics of $\rho$. 

Note that the anomalous behavior has the following universality property: 
 if the exponent $\alpha_\Psi(\rho)$ is greater than $\frac{1}{2}$, then the asymptotic behavior of $\Psi_N(\rho)$ is determined only by the tails of $\rho$. Indeed, the contribution of any finite $B\subset\K$ to the maximal variation is of the order of $\sqrt{N}$ and, hence, can be neglected. This explains why further investigation of heavy-tailed setting can be even simpler than investigation of the classical setting with finite $\K$: for finite $\K$ one should take into account the contribution of each atom of $\rho$.  

Another question for the further study is to characterize the slowest speed of error term decreasing
for games with uncountable $\K$. To avoid pathological situations in this case (as in the end of Section~\ref{sect_game_with_anomalous}) some regularity assumptions on one-stage payoffs should be made. Therefore, the problem is to find the impact of regularity on behavior of the error term. An approach allowing to take regularity into account in upper estimates on the error term is developed by F.~Gensbittel in~\cite{Gensbittel2013_fresh}.
The main ingredient of this approach is the maximal variation, where instead of the total variation norm the Kantorovich (Wasserstein) metric is used.

\section*{Acknowledgments}
I am deeply indebted to V.~Domansky, V.~Kreps, and 
E.~Presman for constant attention to this work and for their support.
I grateful to B.~De~Meyer, F.~Gensbittel, I.~Ibragimov, A.~Neyman, N.~Smorodina, and S.~Zamir for encouraging discussions and useful remarks. I also would like to thank two anonymous 
referees for their helpful comments and suggestions.

The research is supported by the Chebyshev Laboratory
(Department of Mathematics and Mechanics, St.~Petersburg State University)
under the Russian Federation Government grant 11.G34.31.0026, by JSC ``Gazprom Neft'', and by the grants 13-01-00462 and 13-01-00784 of
the Russian Foundation for Basic Research.


\begin{thebibliography}{}

\bibitem{A-M}
Aumann R, Maschler M
(1995)
Repeated games with incomplete information.
MIT~Press, Cambridge 

\bibitem{Azuma}
Azuma K
(1967)
Weighted Sums of Certain Dependent Random Variables.
Tohoku~Math~J 19:357-367

\bibitem{DeM1996_1} De Meyer B 
(1996) 
Repeated games and partial differential equations. 
Math Oper Res 21(1):209-236

\bibitem{DeM1996_2} De Meyer B 
(1996) 
Repeated games, duality and the central limit theorem. 
Math Oper Res 21(1):237-251

\bibitem{DeM1998} De Meyer B 
(1998)
The maximal variation of a bounded martingale and
the central limit theorem.
Annales de l'Institut H Poincare (B) Probability
and Statistics 34(1):49-59

\bibitem{DeM2010} De Meyer B
(2010)
Price dynamics on a stock market with asymmetric information.
{Games Econ Behav 69:42-71}

\bibitem{CavU} Gensbittel F
(2012)
Extensions of the $\cav(u)$ theorem for repeated games with incomplete information on one side.
HAL preprint 
http://hal.archives-ouvertes.fr/hal-00745575. Accessed 01 September 2013

\bibitem{Gensbittel2013} Gensbittel F
(2013)
Covariance control problems of martingales arising from game theory.
SIAM J. Control Optim 51(2):1152–1185

\bibitem{Gensbittel2013_fresh} Gensbittel F
(2013)
Continuous-time limit of dynamic games with incomplete information and a more informed player.
HAL preprint 
http://hal.archives-ouvertes.fr/hal-00910970. Accessed 01 December 2013


\bibitem{MZBothSides}  Mertens J-F, Zamir S
(1971)
The value of two-person zero-sum repeated games with lack of information on both sides. Int~J~Game Theory 1:39-64

\bibitem{MZNormalGames}  Mertens J-F, Zamir S
(1976)
The normal distribution and repeated games.
Int~J~Game~Theory 4:187-197

\bibitem{MZVariation} Mertens J-F, Zamir S
(1977)
The maximal variation of a bounded martingale.
Israel~J~of~Math 27:252-276

\bibitem{MZ1995} Mertens J-F, Zamir S
(1995)
Incomplete information games and the normal distribution. 
CORE~DP 9520


\bibitem{Bigbook} Mertens J-F, Sorin S, Zamir S
(1994)
Repeated Games.
CORE~DP 9420, 9421, 9422.  CORE, Louvain-La-Neuve

\bibitem{Neyman} Neyman A
(2012)
The maximal variation of martingales of
probabilities and repeated games with incomplete information.
J.~Theor Probab 26:557-567

\bibitem{Fedor_doklady}Sandomirskii F
(2012)
Variation of martingales taking their values in probability measures and
repeated games with incomplete information.
Doklady Mathematics 86:796-798

\bibitem{Srivastava} Srivastava S M
(1998)
A Course on Borel Sets (Vol. 180). 
Springer 

\bibitem{Zamir_sqrt_is_precise} Zamir S
(1971)
On the relation between finitely and infinitely repeated games with incomplete information.
Int~J~Game Theory 1:179-198
\end{thebibliography}
\end{document}